\documentclass[12pt]{article}
\usepackage[margin=30mm]{geometry}
\usepackage[utf8]{inputenc}
\usepackage[english]{babel}
\usepackage{authblk}
\usepackage{amsmath}
\usepackage{amsfonts}
\usepackage{amssymb}
\usepackage{graphicx}
\usepackage{verbatim}
\usepackage{float}
\usepackage{url}
\usepackage{caption}
\usepackage{subcaption}
\usepackage{rotating}
\usepackage{natbib}

\usepackage{array}
\newcolumntype{P}[1]{>{\raggedright\arraybackslash}p{#1}}

\begin{document}
\title{Network Sampling Methods for Estimating Social Networks, Population Percentages, and Totals of People Experiencing Unsheltered Homelessness}

\author[1]{Zack W. Almquist\thanks{Corresponding Author: zalmquist@uw.edu}}
\author[2]{Ashley Hazel}
\author[3]{Owen Kajfasz}
\author[3]{Janelle Rothfolk}
\author[3]{Claire Guilmette}
\author[4]{Mary-Catherine Anderson}
\author[1]{Larisa Ozeryansky}
\author[1]{Amy Hagopian}

\affil[1]{University of Washington}
\affil[2]{University of California, San Francisco}
\affil[3]{King County Regional Homelessness Authority}
\affil[4]{Stanford University}

\maketitle
\clearpage
\begin{abstract}
\noindent
In this article, we propose using network-based sampling strategies to estimate the number of unsheltered people experiencing homelessness within a given administrative service unit, known as a Continuum of Care. We demonstrate the effectiveness of network sampling methods to solve this problem. Here, we focus on Respondent Driven Sampling (RDS), which has been shown to provide unbiased or low-biased estimates of totals and proportions for hard-to-reach populations in contexts where a sampling frame (e.g., housing addresses) is not available. To make the RDS estimator work for estimating the total number of people living unsheltered, we introduce a new method that leverages administrative data from the HUD-mandated Homeless Management Information System (HMIS). The HMIS provides high-quality counts and demographics for people experiencing homelessness who sleep in emergency shelters. We then demonstrate this method using network data collected in Nashville, TN, combined with simulation methods to illustrate the efficacy of this approach and introduce a method for performing a power analysis to find the optimal sample size in this setting. We conclude with the RDS unsheltered PIT count conducted by King County Regional Homelessness Authority in 2022 (data publicly available on the HUD website) and perform a comparative analysis between the 2022 RDS estimate of unsheltered people experiencing homelessness and an ARIMA forecast of the visual unsheltered PIT count. Finally, we discuss how this method works for estimating the unsheltered population of people experiencing homelessness and future areas of research.
\end{abstract}

\noindent
\emph{keywords}: homelessness, people experiencing homelessness, social networks, hard-to-reach populations, sample data, networks, social networks, respondent driven sampling

\clearpage





\section{Introduction}

The United States (US) Department of Housing and Urban Development (HUD) estimated approximately 582,000 people in the US were experiencing homelessness in HUD's 2022 Homeless Assessment Report to Congress. Counting the number of people experiencing homelessness in the United States is one of the most critical challenges in the demography of the United States. HUD mandates a bi-annual enumeration of sheltered and unsheltered homeless populations by each ``Continuum of Care" administrative unit. Since 2007, this count has been conducted as a one-night ``census" of those people visibly homeless outside of shelters, as well as an administrative count of those registered in shelters during the same period. This process has become known as the ``Point-in-Time" (PIT) survey \citep{almquist_connecting_2020}.

Recently, \cite{tsai2022annual}, and others have critiqued the traditional one-night PIT methodology used by the HUD-designated administrative regions---formally referred to as Continuum of Care agencies (CoC) by HUD. These one-night counts are typically conducted by volunteers trained to look for and tabulate all people sleeping outdoors (i.e., "unsheltered") on that single night. This count is then combined with a shelter count, completed by pulling the records for the same night as the PIT count from the HUD-required Homeless Management Information System (HMIS) system. These combined numbers represent the total homeless population for a given CoC. Aggregating all the CoCs together represents the total number of people experiencing homelessness in the US.

For example, in Nashville/Davidson County CoC, TN, in 2022, it was estimated that 1,916 people were experiencing homelessness on a night in January, but only 634 were estimated to be visibly unsheltered \citep{nashville}. Another example is in King County, WA, wherein in 2020, it was estimated that 11,751 people were experiencing homelessness in a single night in January \citep{hud2020}). However, only 5,578 were estimated to be visibly unsheltered. Further, HUD requires a demographic breakdown of this count (e.g., gender, race/ethnicity, age). For the sheltered population, this information is provided by the HMIS uptake forms; however, for the PIT count, this is typically done with a survey conducted over about a month after the PIT count. This survey is most often conducted as a convenience sample within the CoC. For example, in King County, WA, the 2020 demographics were estimated from a short survey administered at known day site locations, food banks, etc., by a third-party vendor a month after the official PIT count \citep{hud2020} or Nashville/Davidson County CoC which conducts a similar PIT count in January. 
Nationally, the California-based ``Applied Survey Research" firm has advised cities around the country for 40 years on conducting the ``one-night count" using its methods.\footnote{\url{https://www.appliedsurveyresearch.org/}}

Even though the PIT count has come to dominate the discussion of how people experiencing homelessness are counted, there is a long history of proposed methods for counting people experiencing homelessness in the literature (see \cite{roncarati2021invited,snow1991researching} for example). However, recent developments in network sampling and movements by CoC leadership as well as qualitative studies \citep{snow1991researching} provide a strong argument for employing some form of network sampling as a general method for counting the homeless population and subgroups of interest (e.g., LGBTQ+ populations) as one way to handle the critiques leveled by \cite{tsai2022annual} and others instead of the current census-based methodology. This format could take a number of potential frameworks; however, the most robustly developed method for sampling without a sampling frame is probably respondent-driven sampling (RDS), a sampling procedure that attempts to reproduce a random walk over the target population's social network \citep{abdesselam2020development,johnston2016systematic,heckathorn2017network}. Typically, this is done by first obtaining a small convenience sample of people in the target population (e.g., people experiencing homelessness) and providing each person (known as a ``seed") with a ticket to recruit someone from their network who is also in the target population. Tickets are distributed to each person interviewed (so they can invite their friends and contacts to be interviewed) until a sufficient trace into the network has been completed. This resultant sample can be reweighted to provide an unbiased estimate of the population proportion \citep{gile20107}.

In this paper, we lay out a strategy for conducting a network sample of unsheltered people (combined with HMIS administrative data of sheltered but unhoused people) to obtain cost-effective and high-quality estimates of people experiencing homelessness at a ``point-in-time" (which can be over a month or so) in a CoC. We lay out the formal development of the proposed methodology---an RDS approach that leverages known administrative data---and demonstrate the effectiveness of this approach with a network study of people experiencing homelessness in Nashville, TN. Then, we use these data to illustrate how we can estimate complete networks from our sampled data and how that could be used to generate a simulated RDS for power analysis (i.e., set the level of precision we need for an RDS study based on policy needs). Then, we review the use of this method for estimating the total number of unsheltered people experiencing homelessness in King County, WA, in 2022 and compare it against a forecasted estimate of the visual PIT count for comparison. Finally, we conclude with a discussion of applications of this approach for CoCs and potential further developments.

\section{Counting People Experiencing Homelessness}

\paragraph{History of governmental counting of homelessness:} Counting people experiencing homelessness in the United States began in the 1970s and 80s. In the 1980s, homelessness in the US came under the national spotlight when advocates---e.g., the National Coalition for the Homeless---described it as an epidemic encompassing millions of people \cite[e.g.,][]{hombs1986homelessness}. This emerging political relevance engendered a sense of contention that motivated a debate on how to measure the homeless population in the US. These assertions further pushed the US federal government into action to acquire high-quality counts of the homeless population. 

In 1983 and 1984, the Department of Housing and Urban Development (HUD) conducted a series of surveys \citep{bobo1984report,burt1989differences} that were later followed up by two major studies: (1) one by the US Department of Agriculture (USDA)\footnote{USDA was interested in issues around the rural US and feeding people experiencing homelessness \citep{burt1988feeding}.}, which funded a study in 1987 to enumerate a national sample of unsheltered people in the US, and (2) the prototype method for systematic local-enumerations based on stratified samples of microgeographies, which was piloted in the city of Chicago. The USDA study attempted to count the number of people experiencing homelessness and record their demographics in 20 major cities across the US. It is considered the first nationally representative dataset for homelessness in the US \citep{poulin2008history}. The systematic local enumeration method, pioneered in Chicago, was conducted and written up by \citet{rossi1987urban} and greatly influenced modern methods for estimating the size and composition of homeless populations. 

The 1987 McKinney Act tasked the US Census Bureau with counting the homeless population during the regularly scheduled census; the largest share of this responsibility (per the McKinney-Vento Act) fell to the US Department of Housing and  Urban  Development \citep{lee-anderson_is_2019}. The US Census Bureau thus followed up the Chicago study as part of its 1990 census, during which an intensive study was conducted of five major US cities---collectively known as the S-Night sample  \cite[where ``S" stands for both street and shelter;][]{barrett19901990}. This experiment, run by the US Census Bureau, is the basis for the modern-day point-in-time count by HUD. 

In 1999, Congress directed HUD to create a vehicle for national reporting of homeless counts. This requirement would eventually form the Continuum of Care area units/communities started in 2005, and PIT counts conducted since 2007 \citep{coc101}. Thus, HUD’s Annual Homelessness Assessment Report has become the authoritative estimate of the homeless population size. The HUD counting methodology for the unhoused population includes estimates on a given night (i.e., the PIT count) and estimates over the course of a year, known as ``annual prevalence" \citep{metraux_posthumously_2016}. 

\paragraph{Critiques of governmental counting of people experiencing homelessness:}
 The primary critique of the PIT method pertains to its propensity to undercount homeless populations. This happens because individuals must be visible to count them, environmental conditions impede visibility, the desire of individuals to remain hidden, HUD’s restrictive definition of homelessness, and variability in enumerator counting methodology. There have also been critiques of the changes to inclusion criteria such as classification of transitional and permanent housing services that can make the interpretation of these statistics opaque \citep{hopper_estimating_2008,institute_of_medicine,lee-anderson_is_2019,noauthor_dont_2017,schneider2016we}. The subjectivity, bias, and precarity of PIT estimates have been shown to affect the counts \citep{golinelli_strategies_2015} significantly. For instance, New York’s Department of Homeless Services interviews visible homeless people between midnight and 4 AM on a given night each year and has conducted experiments to estimate the number of people missing in a given year. This has resulted in an estimate of 30\%–40\% of unsheltered people experiencing homelessness being in a location that was hard to see by the PIT surveyors \citep{hopper_estimating_2008}.

 Advocates have increasingly pointed out discrepancies between the PIT-count estimates and suggested reality. They maintain that conventional estimates minimize the problem and lead to misinformation and insufficient response to the homeless crisis in the United States. Much evidence supports their claims: for instance, a general population sampling approach conducted in Los Angeles found that adding the unsheltered homeless persons who were missed in a PIT enumeration increased the estimated size of the total population by more than 20\% \citep{agans_enumerating_2014}. 

\paragraph{Academic research studies estimating the number of people experiencing homelessness:} Several one-off small area estimation case studies of homeless populations have been conducted in major US cities (e.g., Los Angeles; \citet{berk2008counting}), as well as intensive surveys of people experiencing homelessness (e.g., US Census S-Night count). \citet{berk2008counting} uses the pre-{Continuum of Care} mandated point-in-time count for LA in 2004 to build estimates of county, city, and census-tract level homelessness for Los Angeles County. The researchers sampled data for census tracts at two waves and employed a random forest approach to modeling and aggregating tract sample data (totals are estimated using a Horvitz-Thompson estimator). Others have used the US Census S-Night sample to understand the measurement of homeless populations and make small area estimates \cite[e.g.,][]{lee2004geography,bentley1995measuring,wright1992counting}.

A handful of longitudinal studies exist, such as \citet{link1995lifetime}'s study of five-year prevalence rates of homelessness. Given the prevalence of the homelessness problem, surprisingly limited use has been made of the nationwide homeless count data (available from HUD for academic research). This is because it's not linked to major social science datasets like the US Census. Notable exceptions include \citet{byrne2013new} and subsequent follow-up research on community factor predictors of homelessness. However, the only significant collection of homeless count data is the US Department of Housing and Urban Development's Continuum of Care Census count. More recently, \cite{agans2014enumerating} suggested a phone-based survey approach to improving the standard PIT, and \cite{bergmann2021strategies} in a policy report to Hennepin County, MN, suggested a spatial sampling approach for future PIT counts.

\paragraph{Advocacy and NGO research studies estimating the number of people experiencing homelessness:} Significant variations in the counting methods are employed by research organizations, advocacy organizations, and communities to count homeless populations (PIT-style methods are also required for HUD Continuum of Care funding). These estimates include posthumous assessments, recorded uptake of shelter and homelessness services, location sampling, and innovative approaches like ``plant-capture strategies," among others \citep{hopper_estimating_2008,golinelli2015strategies}. A common criticism of the PIT count by advocates is that it was started locally as a non-profit-led community effort to bring awareness to the problem of people experiencing homelessness rather than a rigorous count of the unsheltered population. For example, in King County, it was originally organized locally by the Seattle King County Coalition on Homelessness (SKCCH) as a community-building exercise.\footnote{\url{https://homelessinfo.org/}}

\paragraph{Discussion on census versus sampling for policy decisions:} The Supreme Court case of ``Department of Commerce v. United States House of Representatives" in 1999 addressed the use of statistical sampling in the US census. The case questioned whether the Census Bureau could employ sampling instead of a complete headcount to estimate population. In a 5-4 decision, the Court ruled against statistical sampling for apportionment, stating it violated the constitutional requirement of an ``actual enumeration." This decision set a precedent, affirming that a census for apportionment must involve a direct headcount, influencing subsequent census methodologies, including the 2000 Census. Because of this historical ruling, people experiencing homelessness in the US census are counted through a direct head count by going to outdoor locations (e.g., woods or parks), emergency and transitional housing, soup kitchens, and regularly scheduled mobile food vans \citep{USCensus2020hc}. Critiques of this \citep{meyer2023size} demonstrated that the US census microdata included a number of duplicate entries. Further, \cite{meyer2023size} showed that around 90\% of those in shelters appear in the US Census. However, the unsheltered PIT count is conducted much more like the American Community Survey (ACS) or the Current Population Survey to supplement the HMIS data (discussed in Section 5.1) than the actual US decennial census as its use is for yearly or bi-yearly estimates. This is similar to the ACS, which provides yearly demographic estimates for the US rather than ten-year estimates for drawing political boundaries. HUD further emphasizes this by providing KCRHA with special method exceptions for King County's unsheltered PIT count through RDS methods; further, while RDS is a special method that may require an exception, there is significant precedent for non-census methods (i.e., sample-based methods) in the PIT count data \citep{HUDMethod}.

\section{Alternative Methods for Unsheltered PIT Counts}

We present four types of major alternatives to the visual census unsheltered PIT count: (1) databases, passive data collection, and engaging the service community; (2) combining survey and administrative data; (3) improving the visual unsheltered PIT count through technology; and (4) online surveys and network methods. We discuss each below, and we include a table summarizing each of the categories and comparing it to the RDS method we propose in this article (see Table~\ref{tab:methodsRds}).\footnote{\cite{bergmann2021strategies} also has an excellent summary of recent PIT strategies in the context of Hennepin County CoC, MN.}

\paragraph{Databases, passive data collection, and engaging the service community:}\cite{tsai2022annual} propose two leading solutions to the visual census PIT count:

(1) Continuous Monitoring and Technology Integration: Implementing a nearly real-time monitoring system using technology, such as mobile apps and data analytics, could provide continuous, comprehensive enumeration data. This real-time data analytic strategy would help overcome the limitations of a one-day count and capture fluctuations over time. Passive monitoring poses significant limitations, however. One is methodological: Many people do not have cell phones, or they lose them or share them or turn them off to preserve battery, and disambiguation between the unhoused population, service workers, and the housed population, who are often all in the exact location, can be an almost intractable problem. Another major issue with this process is the matter of consent; people generally object to being surveilled without their knowledge. The authors conducted a series of focus groups with people experiencing homelessness in King County in 2022 to ask about this issue, and it was abundantly clear unhoused people do \emph{not} like the idea of monitoring cell phone signals to obtain counts of people experiencing homelessness. People in our focus groups specifically called it ``Big Brother-esque" monitoring.\footnote{Approved by the University of Washington IRB, 2022 (Study ID: STUDY00015310).} 

(2) Collaboration and Community Engagement: Involving local communities, social service organizations, and volunteers in the data collection can enhance the count's accuracy. This process is widespread in street counts, as an organization needs many people to cover a given Continuum of Care (historically, King County had around 1,000 volunteers). This approach may help identify and reach homeless individuals not easily accessible during traditional counts. This collaboration and community engagement built through this lively annual event served a purpose, but required significant and expensive organizational collaboration. Still, the collateral benefits were important, and could still be integrated into the RDS strategy.

\cite{ward2022recent}, while not advocating their strategy as a method for replacing the PIT count, do provide a framework for a spatial skip sampling procedure (randomly sampling spatial areas and then interviewing every other or every fifth person experiencing homelessness that the interviewer comes across). This method provides estimates of the demographic proportions of people experiencing unshelteredhomelessness and should provide very high-quality proportion estimates. One could adapt the estimator in Section 7 from this type of data to also produce a total estimate of people experiencing homelessness. Two significant limitations are worrisome: 1) it does not provide a hub location format, allowing people to choose to be in the survey or not; the interviewer walks up to the person directly, putting pressure on the respondent to be in the survey---this consent issue is critical---and 2) it is resource intensive because it needs enough staffing to be conducted over a large geographic area in cases like Los Angeles, CA, or even King County, WA. \cite{felix2004combining} lay out a framework for combining short-chain RDS estimators with spatial sampling and would be a natural way to integrate the two processes or supplement either strategy.

\cite{meyer2023size} make an effort to evaluate how well the current Census and ACS cover the population of people experiencing homelessness in the US. They demonstrate that most people covered in emergency shelters in the HMIS data are picked up (around 90\%). However, ultimately, the unsheltered population is estimated using PIT data and thus does not provide an alternative to the visual census count. Other national surveys could also be tried, but we are unaware of any efforts to do so in the US.\footnote{ \cite{o2020estimating} attempts to estimate the Australian national prevalence of people experiencing homelessness from the General Social Survey for Australia; we are unaware of an attempt to do so in the US General Social Survey.}

\paragraph{Survey and administrative data:} The plant-capture/recapture method is an innovative approach borrowed from ecological research and population biology \citep{hopper2008estimating}. In the context of estimating unsheltered homelessness, this method involves placing discrete markers (embedded human decoys) in locations where homeless individuals are known to frequent and then, at a later date, recapturing the same markers. For example, staff members who have received training are tasked with assuming the role of decoys. They are directed to dress and behave as if they are homeless, and afterward, they are expected to provide feedback to the CoC regarding whether they were included in the count that took place on the designated night. An estimate of the unsheltered homeless population can be calculated by determining the rate of marker recapture and using statistical models. This method can improve the estimate of the total unsheltered homeless population but does not provide demographic breakdowns or a proper statistical estimator. 

The capture-recapture method (sometimes called multiple-list methods, see \cite{weare2019counting}) is a statistical approach derived from ecological and wildlife population studies \citep{williams2002can}. To estimate people experiencing unsheltered homelessness using capture-recapture methods involves two or more separate counts of the same population within a defined time frame, allowing for an estimation of the total population size through mathematical modeling. Capture-recapture can address the issue of undercounting in PIT counts by providing a statistical correction. Accurate capture-recapture methods require data from multiple sources, such as outreach teams, service providers, and independent counts, which can be logistically challenging to coordinate. Currently, the method does not provide a way to estimate demographic characteristics, and it typically involves conducting a separate survey rather than just using administrative logs. It could be done simultaneously with an RDS sampling strategy in this context.

Demographic post-PIT surveys involve surveys with a sample of individuals counted during the traditional PIT count. These surveys gather detailed demographic and situational information, shedding light on the characteristics and needs of the unsheltered homeless population. Surveys provide valuable individual-level data, allowing a deeper understanding of the homeless population. These surveys do not directly estimate the total unsheltered homeless population and may not address undercounting issues; however, CoCs have developed questions for these surveys to improve the PIT count. RDS could replace the unsheltered PIT count (such as that done by KCRHA in King County, WA) or as an alternative to the demographic survey and thus provide a second count of the total number of unsheltered people experiencing homelessness. 

\paragraph{Visual unsheltered PIT technology improvements:} GIS tools offer a robust set of resources to improve the accuracy and value of the unsheltered PIT count. Incorporating spatial analysis, geocoding, mobile data collection, and integrated data recording with classic visual unsheltered PIT count can improve the overall count process. The addition of apps and automated geocoding has been used to improve the accuracy and reliability of classic unsheltered PIT count; unfortunately, this expensive approach requires training for volunteers and does not provide statistical reliability estimates or demographics, so it has to be paired with post-PIT demographic surveys. 

\paragraph{Online surveys and network methods:} Network scale-up methods (NSUM; \cite{laga2021thirty}) have been used in the academic and public-health literature to estimate people experiencing homelessness \citep{killworth1998estimation}. This method works by taking a random sample from a known population (e.g., landline, cellphone, or address-based sample) and asking respondents how many people they know in each category, including the priority population of people experiencing unsheltered homelessness. Recent improvements in this method could be employed to better understand the unsheltered population (see, for example, the work of \cite{feehan2016generalizing}). Also, scale-up methods have been employed with online surveys to estimate offline populations using a random sample of Facebook users \citep{feehan2019using}; this strategy has the potential for estimating the unsheltered population in the US. The NSUM methods should, in principle, be adapted to the RDS design proposed in this work by adapting \citep{breza2023consistently} network data aggregation methods that could be used to provide another quality check on the RDS method when combined with NSUM methods. 

Last, \cite{maas2020using,giraudy2021measuring} demonstrates how Facebook can be used to sample displaced people post-disaster in Australia, including comparable estimates to the UN's for total people displaced during a bushfire. This method could also be adapted to estimate the total number of unsheltered people experiencing homelessness. 







\begin{sidewaystable}
    \centering
    \footnotesize
    \textbf{Comparison of alternative methods for counting people experiencing unsheltered homelessness}\\
   \begin{tabular}{P{2.5cm}|p{9.5cm}|p{9.5cm}}
Method & Description & Relation to RDS \\
\hline
\hline
\multicolumn{3}{c}{Databases, passive data collection, and the service community}\\
\hline
HMIS databases & Centralized databases at the CoC level cover most service activities. It does not collect data specifically on those living unsheltered or not using services.& Similar to other methods in the list below, HMIS provides the administrative backbone needed to estimate homeless populations for the RDS method.\\
 \hline
Census data linkage & The American Community Survey and Decennial Census collect information on people experiencing homelessness and can be linked to HMIS. Coverage of the population of people using shelters is high (around 90\%) but does not cover the people experiencing homelessness who live outside the shelter system. & This method is not robust for counting the unsheltered population and is not a considered as an alternative for the proposed RDS method.\\
\hline
Big data/cellphones & Uses cellphone and app data such as that collected by Safegraph or other data aggregators.  & Difficult to implement and potentially viewed as unethical for establishing the population size of people experiencing unsheltered homelessness.\\
 \hline
 Survey of service workers & Surveying survey works post unsheltered PIT count to improve final count.& This could be done post-RDS survey too. \\
\hline
\multicolumn{3}{c}{Survey and administrative data}\\
\hline
Plant-capture/recapture methods & Uses evenly dispersed human decoys to estimate how many people are missed in the visual unsheltered PIT count.& Alternative way to improve PIT count does not provide explicit statistical bounds like RDS.\\
\hline
Capture-recapture (multiple-list) methods & Uses a statistical approach to estimate the total population size by comparing differences between two or more lists.& This can be done with administrative data (e.g., service lists compared to HMIS) or surveys (e.g., post-demographic survey or RDS). It is straightforward to implement alongside RDS to give an alternative estimate to the RDS for quality purposes. However, it needs to provide a way to estimate demographic breakdowns (this is a current research problem in the capture-recapture space).\\
\hline
Post unsheltered PIT count survey & Survey conducted after unsheltered PIT count to estimate demographic percentages. It can be used to estimate the number of people who were missed. & Alternative to RDS or the RDS method proposed could replace the post-unsheltered PIT demographic survey (do both methods). \\
\hline
\multicolumn{3}{c}{Visual unsheltered PIT technology improvements}\\
\hline
App GIS tools & App with real-time GPS capabilities used to improve unsheltered visual PIT count.& Should improve unsheltered PIT count, but needs to provide statistical bounds or demographic information. \\
\hline
\multicolumn{3}{c}{Online surveys and network methods}\\
\hline
Network scale-up methods (NSUM) & Standard NSUM methods use landline, cellphone, or address-based random population samples with network questions on how many unsheltered people the person knows to scale up to a population estimate. & Alternative method to RDS, in principle. It is possible to join the RDS and NSUM methods to provide another check on the total population; however, this method has to be adapted to the case where one is sampling the target population rather than the general population.\\
 \hline
Online surveys through marketing platforms & Apps such as Facebook to survey either the general population and combine with NSUM methods or the target population and estimate similar to the method proposed in this paper (similar to methods used in disaster research). & Alternative method for estimating the total number of unsheltered people experiencing homelessness. Research is ongoing.\\
\hline
\end{tabular}
        \caption{\footnotesize Table listing core alternatives to the RDS method proposed here and the basic relationship between the alternative method and the proposed RDS sampling method and estimator. }
        \label{tab:methodsRds}
\end{sidewaystable}

\section{Snowball, Respondent-Driven Sampling, and Other Network Approaches}

A general design challenge in survey research is how to sample a population when there is no sampling frame available (e.g., a list of people experiencing homelessness) or the population is a minimal subset of the total population (e.g., Indigenous populations). Researchers have approached this problem in several novel ways depending on resources and available information. Suppose we have access to the networks of a population of interest (where a network is a collection of individuals and their social relations, such as friends or acquaintances). In that case, one solution is to sample the population of interest by leveraging social ties between individuals. This requires some assumptions, the key of which is that the population of interest is well connected (e.g., that people experiencing homelessness know other people experiencing homelessness). If this is true, we can employ \emph{network sampling} methods to obtain an unbiased (or minimally biased) estimate of the target population. In its simplest form, a network sample replicates a mathematical random walk through the focal population (e.g., people experiencing homelessness) where, at each step, the location jumps to another site (or person) according to a probability distribution (e.g., uniformly weighted by the number of alters an individual has). In a simple random walk, the location can jump only to neighboring nodes (or people) of the network (e.g., an individual's friendship or acquaintance network), forming a lattice path. Such a process can be reweighted to produce an unbiased estimator via Horvitz–Thompson (sampling without replacement) or Hansen-Hurwitz (sampling with replacement; see, for example, \cite{salehi2003comparison}). This procedure (as well as the Metropolis-Hastings algorithm) is possible in online settings, and details of these estimators have been explored in \cite{gjoka2010walking}. However, it is often not possible for in-person settings to do a proper statistical random walk through the network. First, typically, the design does not allow for sampling with replacement, and second, we usually have limited ability to randomize node selection in a respondent's personal network. 

Two sampling designs routinely used in social science and public health research are \emph{snowball} sampling and \emph{respondent-driven sampling} (RDS). Snowball sampling (where respondents supply names of their known contacts) is typically used for either ``obtaining a nonprobability sample through an unspecified network search process or constructing a frame from which to sample" \citep{handcock2011comment}. In either case, it is not formally the same process as RDS, which is, as \cite{handcock2011comment} describes, a link-based sampling approach. In other words, RDS is a process for formally sampling (i.e., randomizing) the social network rather than a purely convenience sample of the target population. Further, university IRBs prefer this method because it allows the referred person to decide whether to come to the researcher rather than the researcher pursuing the subject. For improved sampling, the seeds can be randomized, e.g., spatial sampling, which is a strategy that is likely to be very effective for sampling people experiencing homelessness.
In general, RDS is a method for both data collection and statistical inference that relies on selecting a small number of initial participants (``seeds") from the target population who are asked to recruit their contacts in the target population (usually, this is followed with incentives) and includes a randomization element. This procedure is repeated until a desired sample size is reached. The typical recruitment size is three per a given contact's personal network. Because this process is known to over-sample people with many connections, the sample is reweighted to provide an unbiased (or minimally biased) estimate. 

\begin{table}[H]

    \textbf{Original \cite{volz2008probability} RDS Assumptions}\\
\begin{tabular}{p{6cm}|p{8cm}}
  \textbf{Sampling Assumptions}   & \textbf{Population/Network Assumptions}  \\
   \hline
 The seeds are chosen with probability proportional to their network degree & Potential recruitment is symmetric (individual i would list person j as a potential recruit if j would list i)\\
\hline
 All respondents receive and use only one coupon & The network is connected such that there is a path from each individual to every other individual\\
 \hline
Each recruiter chooses their recruit at random from among their
contacts
\end{tabular}
    \caption{\cite{volz2008probability} assumptions for the original Respondent Driven Sampling estimator.}
    \label{tab:rds}

    \textbf{\cite{gile2015diagnostics} RDS Assumptions}\\
\begin{tabular}{p{3cm}|p{5cm}|p{5cm}}
& \textbf{Network assumptions}	& \textbf{Sampling assumptions}\\
\hline
Random-walk model	& Network size large ($N\leq n$)& With-replacement sampling Single non-branching chain\\
\hline
Remove seed dependence & 	Homophily sufficiently weak, Bottlenecks limited and Connected graph &	Enough sample waves  \\
\hline
Respondent behavior & All ties reciprocated  &	Degree accurately measured and  Random referral\\
\end{tabular}
    \caption{\cite{gile2015diagnostics} updated assumptions for the original Respondent Driven Sampling estimator.}
    \label{tab:rds2}
\end{table}

\cite{volz2008probability}, in the original paper, provided an essential set of assumptions for respondent-driven sampling so that its idealized process would result in a random walk through the social network of the target population. The exact table can be seen in Table~\ref{tab:rds} and \cite{gile2015diagnostics} propose an alternative set of assumptions in Table~\ref{tab:rds2}. Under these assumptions, each tie has an equal probability of being sampled. Therefore, the probability that an individual is included in the sample is proportional to the degree ($\pi_i=cd_i$, where $c$ is an unknown constant and $d_i$ is the network degree (number of contacts in the target population) for individual $i$). \cite{volz2008probability} shows that the sampling assumption that ``the seeds are chosen with probability proportional to their network degree" can be removed given that the recruitment chains are long enough such that the probability of sampling each individual at the next step of the chain has reached a Markov equilibrium, and further that all observations before achieving equilibrium are ignored. This results in what is known as the \emph{Hajek} estimator \cite[see for example][]{fellows2022robustness}. Several papers have evaluated the robustness of these estimators and proposed alternatives \cite[e.g.][]{gile2011improved,gile2015diagnostics}.

\begin{figure}[H]
     \centering
     \begin{subfigure}[b]{0.4\textwidth}
         \centering
         \includegraphics[width=1\linewidth]{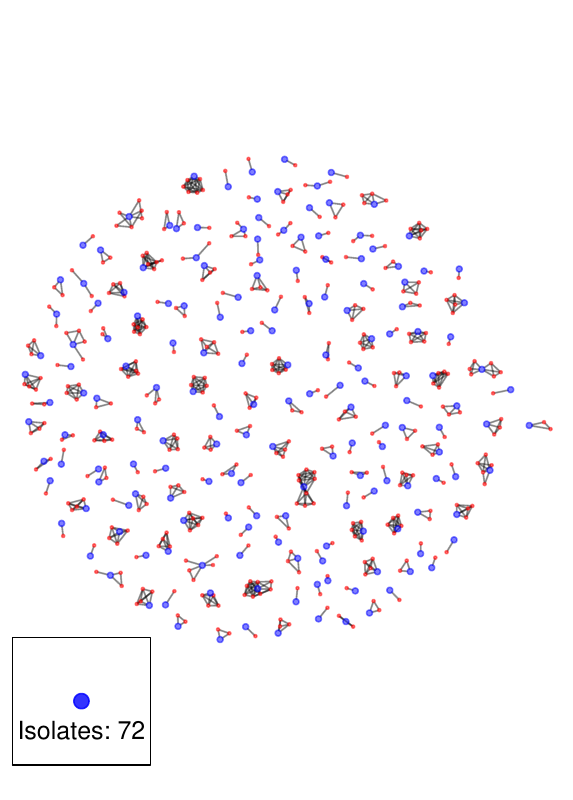}
    \caption{A census of egonets (personal networks) for \cite{anderson_ecology_2021} study in Nashville, TN.}
    \label{fig:egonets}
     \end{subfigure}
     \hfill
     \begin{subfigure}[b]{0.4\textwidth}
         \centering
         \includegraphics[width=1\linewidth]{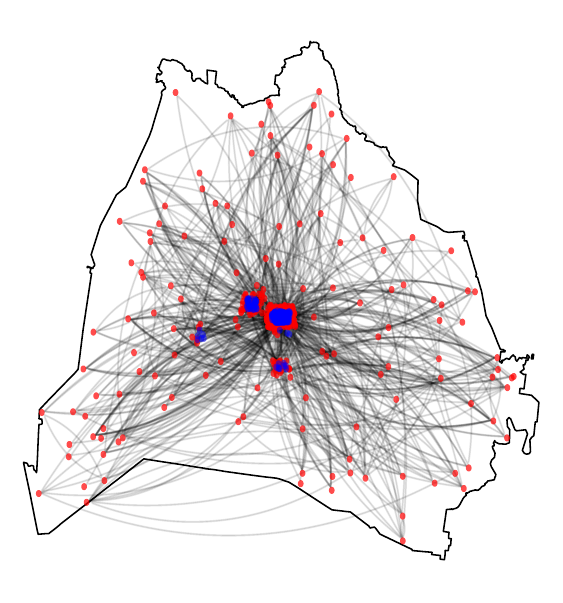}
         \caption{Egonets (personal networks) plotted with spatial embedding for \cite{anderson_ecology_2021} study in Nashville, TN.}
         \label{fig:egonetmap}
     \end{subfigure}
             \caption{Egocentric network plots from \cite{anderson_ecology_2021} study in Nashville, TN. Blue represents surveyed individuals, and red represents reported alter.}
        \label{fig:egonets_plots}
\end{figure}

\section{Data}

This section will review the data and methods for this article. First, we will cover the available administrative data derived from the HUD-mandated Homeless Management Information System (HMIS) databases. Second, we will review the survey data collected over a year in Nashville, TN, from people experiencing homelessness.


\subsection{Administrative Data}

\paragraph{Continuum of Care:} In 1994, the US Department of Housing and Urban Development began requiring each ``community" to come together to submit a comprehensive Continuum of Care application rather than allowing applications from individual providers in a community. These local coalitions provide stable administration for US homeless services and an incentive structure to centralize homelessness planning in specified areas. HUD has been working with local CoC agencies to give estimates of homeless persons in the nation since 2007. Each CoC adapts HUD's standard approach, the homeless PIT count, to estimate the number of homeless people in its jurisdiction.

\paragraph{Homeless Management Information System:} HUD requires CoCs to follow data collection and administrative control standards for its HMIS database systems \citep{culhane2004homeless}. The HMIS can provide a count of unhoused people who engage with the federal reporting system. Specifically, HUD provides a series of data standards, which it updates yearly through the HMIS data standards manual. The HMIS Data Standards Manual is a resource that assists HMIS leads/system administrators, Continuums of Care (CoCs), and HMIS end users in their data collection and reporting endeavors. It guides the essential data elements needed in an HMIS to ensure compliance with HUD and federal partners' participation and reporting mandates \citep{HUDHMIS}. Every CoC maintains an HMIS database, which conforms to this set of standards. 


\subsection{Survey Data}

\citet{anderson2021characterizing} collected egocentric network, demographic, economic, and spatial data for her dissertation entitled \emph{Social Networks, Environment, and Well-Being: A Case Study Among People Experiencing Homelessness (PEH) in Nashville, TN}. Dr. Anderson collected 246 egocentric networks (see Figure~\ref{fig:egonets}) and complete alter (e.g., friend) information, demographics, health, psychology, and economic information over four time periods \citep{anderson_ecology_2021}. From June 2017 to October 2017 and December 2018 to February 2018, Dr. Anderson conducted ethnographic and participant-observation fieldwork in Nashville, TN, in informal settings to build rapport with unhoused people and service providers. Dr. Anderson developed a survey administered to participants, eliciting information on sociodemographic characteristics, homelessness history, health, social support and social network composition, and environmental exposure. She piloted the survey by recruiting ten unhoused individuals (compensated for their time) who talked through how certain aspects of the survey, such as language, could be improved. 

From August 2018 to June 2019 in Nashville, TN, survey data were collected from 246 people experiencing homelessness via convenience sampling at four brick-and-mortar service facilities and three street locales where homeless services, such as mobile meals and shower trucks, were offered one day a week. These survey sites were located in various areas in Nashville and are displayed in Figure~\ref{fig:egonetmap}. Unhoused individuals were recruited at each sample site. The research team collected data using paper rather than electronic surveys to be as transparent as possible and to avoid suspicion raised by electronic devices. Following survey completion, participants were given a \$15 gift card.



\section{Estimating the Total Number of Unsheltered People Experiencing Homelessness}

\paragraph{The Problem:} A variety of strategies can be employed to estimate unsheltered homelessness population numbers and proportions (e.g., percent female) from survey data. Among the most common is leveraging known information to estimate population size for the subgroup of interest and applying likely proportions. In this study, we employed network sampling without a sampling frame, combined with administrative data, to obtain estimates of the population size of unsheltered people experiencing homelessness.

\paragraph{Network Sampling:} The primary use of respondent-driven sampling and other link-trace designs is to estimate the population proportion of the target population. For example, what is the proportion of people experiencing homelessness who are unsheltered? 

Recent work has shown the standard RDS estimator is robust to measurement error and is also robust to most sampling assumptions (see \cite{fellows2022robustness}). However, in the context of people experiencing homelessness, two major questions pertain: (1) How robust is the estimator to these sampling assumptions, and (2) can we estimate the total number of the unsheltered population and HUD-required demographics? 

First, we review the use of RDS and its resultant estimators for a ``hypothetical" PIT count where we leverage known administrative data to go from proportions to counts.

\subsection{Totals and Proportions}

Leveraging the RDS estimator, built from the classic generalized Horvitz-Thompson estimator \cite{gile20107} (what \cite{fellows2022robustness} describes as the Psuedo Horvitz-Thompson estimator).

Let us first consider a population of $N$ individuals with a known probability of being sampled, $\pi_i$ of $i$th individual in the sample. We can estimate the population mean, $\mu$, of any quantity, $z_i$, measured on the sampled individuals using a Horvitz-Thompson estimator \citep{gile20107}:
\begin{align}
    \hat{\mu} &= \frac{1}{N} \sum_{i=1}^N S_i \frac{z_i}{\pi_i},
\end{align}
Where $S$ is the random $N$-vector representing the sample, such that $S_i=1$ if unit $i$ is sampled and is otherwise 0. There are two major drawbacks to this estimator in the network sampling context: (1) the population size $N$ is often unknown (in fact, the whole problem is that it is unknown in our case!), and (2) the inclusion probabilities $\pi_i$ are also typically unknown. The first issue can be solved by plugging in the unbiased estimator of $N$, $\hat{N}=\sum_{i=1}^N S_i \frac{1}{\pi}$, to obtain
\begin{align}
    \hat{\mu} &= \frac{\sum_{i=1}^N S_i \frac{z_i}{\pi_i}}{\sum_{i=1}^N S_i \frac{1}{\pi_i}},
\end{align}
which is the ratio of two unbiased estimators and tends to estimate $\mu$ with small bias for large sample sizes. This variant of the Horvitz-Thompson estimator is known as the \emph{generalized Horvitz-Thompson estimator} or the Hajek estimator \citep{gile20107}.

\paragraph{Salganik-Heckathorn Estimator:} \cite{salganik2004sampling} introduced an estimator that leverages the relationship between two groups, $A$ and $B$, where $\bar{d}_A$ is the mean degree of a group $A$, $\bar{d}_B$ is the mean degree of a group $B$, and $c_{AB}$ and $c_{BA}$ are the cross-ties between the two groups. The total for the groups is thus $N_A$, $N_B$ such that $N=N_A+N_B$. Given the above definitions, we can write the following:

\begin{align}
\frac{N_A}{N} = \frac{\bar{d}_B\cdot c_{BA}}{\bar{d}_A \cdot c_{AB}+\bar{d}_B \cdot c_{BA}}=\mu_A
\end{align}

To employ this estimator, we need a way to estimate,

\begin{itemize}
    \item The mean degree of $\bar{d}_A$ and $\bar{d}_B$.
    \item The proportion of relations of group $A$ that are relations to group $B$, and visa versa.
\end{itemize}

Let $t_{AB}$ represent the total number of relations between groups $A$ and $B$. Then, we can write an estimator for the cross ties:
\begin{align*}
    c_{AB} &= \frac{t_{AB}}{N_A\bar{d}_A}\\
    c_{BA} &= \frac{t_{AB}}{N_B \bar{d}_B}
\end{align*}
In practice, \cite{salganik2004sampling} assumed edges were sampled randomly, which allows one to estimate $c_{AB}$ out of the sample proportion of subjects in group $A$ who recruit participants in group $B$. $c_{BA}$ may be estimated in the same way as $c_{AB}$. The mean degrees can be estimated using the Hajek estimator introduced earlier.

If we assume the size of group $B$, $N_B$ is known with little or zero error, we write an estimator of the size of $NA$:

\begin{align} \label{eq:total}
   N_{A}+N_{B} & = N  \\ \nonumber
    N_{B} &= \frac{N_A}{\mu_A} - N_{A} \\ \nonumber
    N_B &= N_A \frac{1-\mu_A}{\mu_A}\\ \nonumber
    N_A &= N_B \frac{\mu_A}{1-\mu_A} 
\end{align}
Notice that $N_B = N_A \frac{\mu_B}{1-\mu_B}$ analogously. We can further estimate standard errors (SE) and confidence intervals through bootstrap methods \citep{baraff2016estimating}. 

\section{Empirical Example: Nashville, TN}

While CoCs currently use a crude census-style approach to counting the number of people visibly experiencing unsheltered homelessness on a single night (along with a count of those in shelters), the current method has a number of limitations and critiques (see \cite{tsai2022annual,roncarati2021invited}). As we advocated earlier, a sample-based approach designed to work in cases where a sampling frame (e.g., household locations) does not exist is a principled solution to a complex problem. While the \citet{anderson2021characterizing} dataset is not a \emph{true} ``respondent-driven sample," it does provide sufficient network data of approximately half the unsheltered individuals experiencing homelessness in Davidson County. We can leverage this dataset to approximate an RDS as a pilot study to see how well we can recreate the PIT data of 2020 in Davidson County, TN.

\begin{figure}[H]
    \centering
    \includegraphics{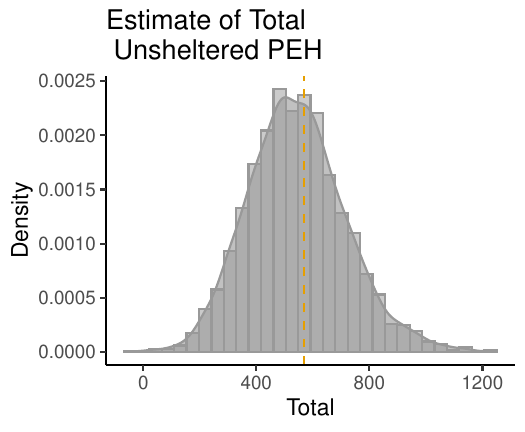}
    \caption{Bootstrap estimates of the unsheltered count of people experiencing homelessness in Nashville, TN, 2020. The red line is the "true" unsheltered count from the Nashville, TN, 2020 PIT survey.}
    \label{fig:bootUScount}
\end{figure}

Building off Equation~\ref{eq:total}, we start by taking $A$ to be the unsheltered population and $B$ to be the sheltered population, then we can rewrite Equation~\ref{eq:total} as,
\begin{align*}
N_u = N_s \frac{\mu_u}{1-\mu_u}.    
\end{align*}
We are assuming that $N_s$ is known. This can be obtained from the HUD HMIS database. The formal PIT count of the sheltered population is typically of high quality and reported publicly in the HUD online database \citep{HUDHIC}. For 2020, the known Nashville population of sheltered people experiencing homelessness was 1,225 \citep{hud2020}. We can estimate $\mu_s$ using the Salganik-Heckathorn estimator\footnote{The Nashville, TN, data are technically a snowball sample and not formal RDS, but there is evidence that applying the same estimator in this context will still result in a low-biased estimator if we are willing to assume the snowball sample is approximating a random breadth-first search process on the network, see \cite{kurant2011towards}.}. This results in $\hat{\mu}_u = 0.304$ (nearly a third of the city's unhoused population was found unsheltered during the point-in-time count). Plugging this estimator into the equation, we find $\hat{N}_u=535$. Further, we employ a bootstrap to produce 95\% CI for this estimate. This results in a mean estimate of 535 with a 95\% CI of [209, 914]. We can compare this result to known unsheltered PIT count data from 2020, which estimated $584$ people \citep{nashville2020}, which is within the 95\% CI. Using bootstrap methods, we can visualize this estimate and the confidence interval in Figure~\ref{fig:bootUScount}.

We can further employ the estimator with a bootstrap confidence interval to estimate the demographic proportions from the estimator in Equation~\ref{eq:total}. It is also essential to explore the estimator's demographic proportion estimates. HUD requires a demographic breakdown (e.g., age or gender) of the PIT account beyond the estimate of the total unsheltered population. In Table~\ref{fig:bootUScountDem}, we see the results of our estimator for the sheltered and unsheltered populations, including the gender and age structure of both groups. We have good coverage (i.e., the 95\% CI covers the ``true" estimate of the population proportion) in all categories, except for the age distribution of the sheltered population. This is likely because the sampling method focused on unsheltered individuals and skewed a bit older in the sheltered population.

\begin{table}[H]
\centering
\begin{tabular}{rrrrrrl}
  \hline
 & Estimate & Lower & Upper & 2020 PIT & Difference & Coverage \\ 
  \hline
Shelter Percent & 0.696 & 0.572 & 0.844 & 0.683 & 0.013 & TRUE \\ 
  Unsheltered Percent & 0.304 & 0.156 & 0.428 & 0.317 & 0.013 & TRUE \\ 
  Gender: Female (Shelter) & 0.323 & 0.235 & 0.426 & 0.275 & 0.047 & TRUE \\ 
  Gender: Male (Shelter) & 0.677 & 0.574 & 0.765 & 0.725 & 0.047 & TRUE \\ 
  Gender: Female (Unsheltered) & 0.323 & 0.225 & 0.421 & 0.233 & 0.090 & TRUE \\ 
  Gender: Male (Unsheltered) & 0.677 & 0.579 & 0.775 & 0.767 & 0.090 & TRUE \\ 
  Age (18–14) Shelter & 0.037 & 0.009 & 0.070 & 0.076 & 0.039 & FALSE \\ 
  Age (Over 24) Shelter & 0.963 & 0.930 & 0.991 & 0.924 & 0.039 & FALSE \\ 
  Age (18–14) Unshelter & 0.057 & 0.014 & 0.103 & 0.029 & 0.028 & TRUE \\ 
  Age (Over 24) Unshelter & 0.963 & 0.897 & 0.986 & 0.971 & 0.008 & TRUE \\ 
   \hline
\end{tabular}
    \caption{Bootstrap estimates of the core demographics of the unsheltered count of people experiencing homelessness in Nashville, TN, 2020.}
    \label{fig:bootUScountDem}
\end{table}

\section{ERGM Estimation and RDS Simulation Using the Nashville, TN, Data}

An important method for the statistical modeling social networks is the so-called Exponential Random Graph Model (ERGM), which allows for the inference and simulation of social networks. Formally, ERGM provides a framework for writing a generative probability model for social networks, see for example \cite{cranmer2011inferential,wasserman1996logit,snijders2006new,hummel2012improving,blackburn2023practical}). In the field of demography, ERGM has been employed to understand racial mixing and other issues \citep{goodreau2009birds}. Given a random graph $G$ on support $\mathcal{G}$, we may write ERGM formally as follows:
\begin{align*}
\Pr(G=g|s,\theta) = \frac{\exp{\theta^Ts(g))}}{\sum_{g^{'} \in \mathcal{G}} \exp(\theta^Ts(g^{'}))} \mathcal{I}_{\mathcal{G}}(g)
\end{align*}
where $\Pr(\cdot)$ is the probability mass of its argument, $\mathcal{G}$ is the support of $G$, g is the realized (observed) graph, and $s$ is the function of sufficient statistics, $\theta$ is a vector of parameters (e.g., the degree distribution of the graph) and $\mathcal{I}$ is an indicator function. This model can be fit by MLE; see \cite{hunter2008ergm} for details. When fit, the results can be displayed in a regression table that can be interpreted in a conditional probability framework similar to logistic regression. 

Suppose we again assume the sample is generated from an approximate random breadth-first search process. In that case, we can re-weight by degree size and use this as the sample weights to estimate an ERG model and simulate the complete network. This process is straightforward for exponential family models, where we build the mean statistics for the model out of sample data and fit with MCMC-MLE. This process was developed in \cite{krivitsky2017inference} and implemented in software \citep{krivitskyEgoNet,hunter2008ergm}. Here, we fit a model that takes into account the population size (this is handled with an offset to the density term), edges (analogous to density), degree up to six, and fixed effects (node factor) for the time period of data collection (fall, winter, spring, and summer), and race/ethnicity and gender. The results are presented in Table~\ref{tab:ergm}.

\begin{table}[H]
\centering
\textbf{ERGM of People Experiencing Homelessness in Nashville, TN, 2020}
\begin{tabular}{l|r|r|r|r|r}
\hline
  & Estimate & Std. error & MCMC \% & z value & $Pr(>|z|)$\\
\hline
offset($N$) & -6.342 & 0.000 & 0 & -Inf & 0.000\\
edges & 2.037 & 1.071 & 0 & 1.901 & 0.057\\
degree2 & 0.338 & 0.491 & 0 & 0.689 & 0.491\\
degree3 & 0.117 & 0.903 & 0 & 0.129 & 0.897\\
degree4 & 0.854 & 1.161 & 0 & 0.735 & 0.462\\
degree5 & 1.300 & 1.144 & 0 & 1.136 & 0.256\\
degree6 & 1.239 & 1.304 & 0 & 0.950 & 0.342\\
$I$(Summer) & -- & -- & -- & -- & --      \\
$I$(Fall)  & -0.881 & 0.841 & 0 & -1.048 & 0.295\\ 
$I$(Winter)  & -0.867 & 0.865 & 0 & -1.003 & 0.316\\ 
$I$(Spring)  & -0.804 & 0.904 & 0 & -0.889 & 0.374\\ 
$I$(White) & -- & -- & -- & --& --      \\
$I$(Black)& -0.016 & 0.433 & 0 & -0.038 & 0.970\\ 
$I$(LatinX) & -0.731 & 0.878 & 0 & -0.833 & 0.405\\ 
$I$(Asian) & 0.308 & 1.482 & 0 & 0.208 & 0.835\\ 
$I$(AIAN) & -0.202 & 0.694 & 0 & -0.291 & 0.771\\ 
$I$(NHPI) & -0.552 & 0.793 & 0 & -0.696 & 0.487\\ 
$I$(Female) & -- & -- & -- & -- & --      \\
$I$(Male) & 0.277 & 0.221 & 0 & 1.251 & 0.211\\ 
Gender Node Match & -0.423 & 0.188 & 0 & -2.246 & 0.025\\
\hline
\end{tabular}
\caption{ERGM parameters fit the Nashville, TN egocentric data. -- represents the reference group. $I$ indicates a node factor, and the "node match" term provides a weight for any time the two individuals have the same gender (known as homophily). }
    \label{tab:ergm}
\end{table}

\section{Simulating Homeless Networks and Power Analysis Strategy for RDS Sampling}

\begin{figure}[H]
    \centering
     \textbf{Simulated Unhoused Network for Nashville, TN, CoC with a Sequence of Simulated RDS Processes}
\includegraphics[width=.9\linewidth]{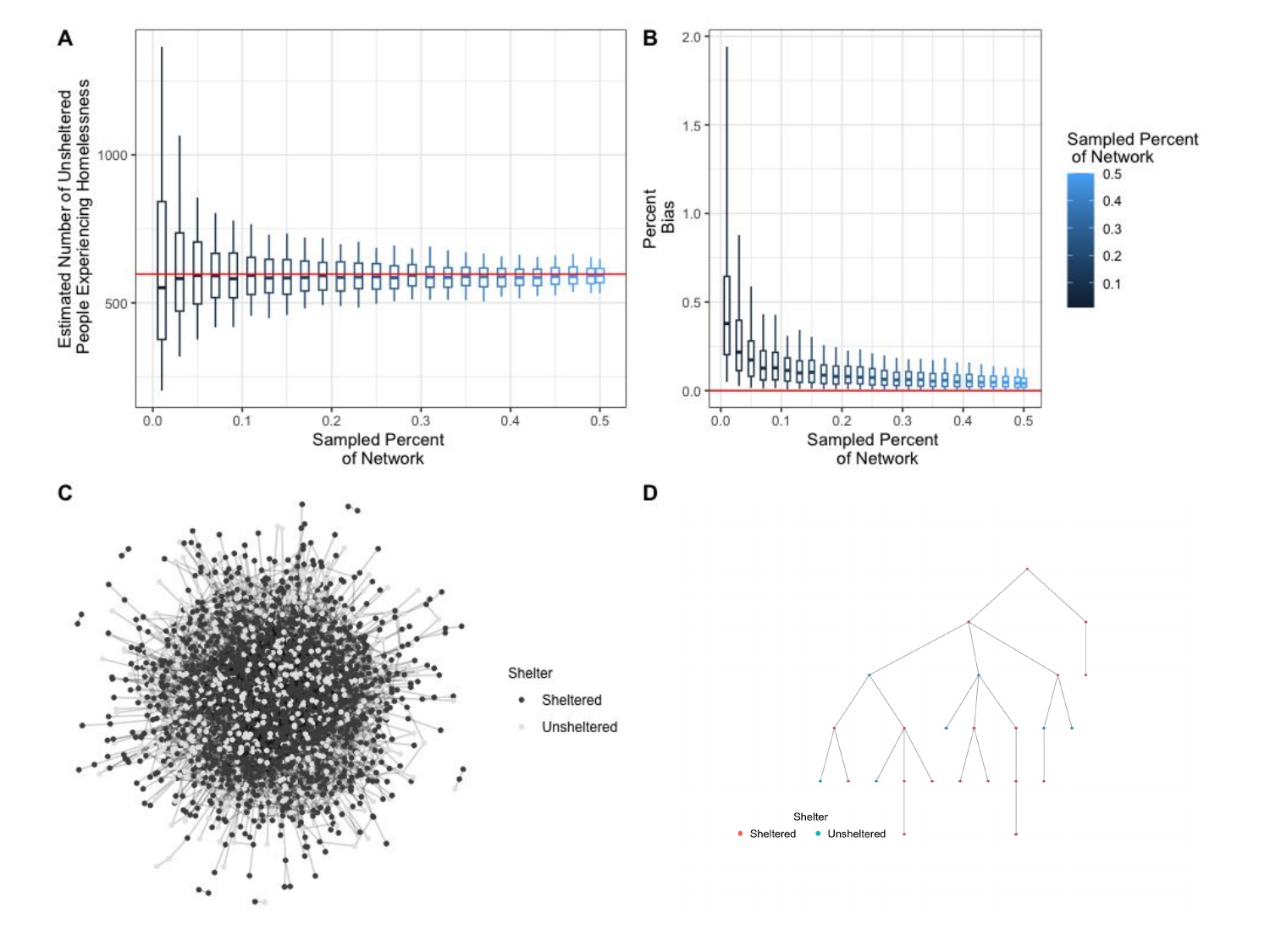}
    \caption{ERGM Simulation of a complete (shelter and non-shelter users) network of people experiencing homelessness in Nashville, TN (2035 people) with 597 unsheltered people and 1,438 people using shelters in the network. \textbf{A} Represents the estimate of the total number of unsheltered people experiencing homelessness compared against the sample size in relative terms (i.e., as a percent of the total population). The red line represents the true value. \textbf{B} Plots the bootstrap estimate of bias against the sample size in relative terms (i.e., as a percent of the total population). \textbf{C} Represents the complete network of 2035 people. \textbf{D} Plot of an example RDS tree.}
    \label{fig:rds_bias}
\end{figure}

We can use this estimation of an ERGM to simulate realistic, complete networks for the unsheltered population of people experiencing homelessness in Nashville, TN, using the same strategy as that employed by \cite{hunter2008ergm} or\cite{almquist2020large}, which specifically simulated networks of people experiencing homelessness. To simulate an RDS process, we can then employ the \cite{rds} R package for generating an RDS process on a social network of people experiencing homelessness. The results of the simulated network can be visualized in Figure~\ref{fig:rds_bias}: C, and the RDS simulation can be visualized in Figure~\ref{fig:rds_bias}:D.

We can then employ this extrapolation ERGM simulation technique to provide a general \emph{power analysis} for an RDS sample for estimating the total number of unsheltered people experiencing homelessness. For example, let's take the case of Davidson County, TN (Greater Nashville, TN). Using our ERGM model fit from the egocentric data, we can extrapolate a complete network of 2,035 with 597 unsheltered people and 1,438 people using shelters in the network. In Figure~\ref{fig:rds_bias}: A, we see the estimated total unsheltered people experiencing homelessness and a bootstrap 95\% confidence interval plotted against the percent of the population sampled. Here, we see that we quickly get about as efficient a sample as we might expect, starting around 0.05\% of a sample and fully stabilizing at around 0.2\% of the sample. Similarly, in Figure~\ref{fig:rds_bias}: B, the sample's statistical bias (sample estimate minus the true value) quickly shrinks with 0.05\% of a sample, fully stabilizing at around 0.2\%. Overall, the statistical bias is quite small in all cases, showing the mean statistic is quite good, although the variance, as is typical with straight RDS, is typically wider than standard survey methods with a sampling frame.



\begin{figure}[H]
    \centering
    \textbf{Simulated Unhoused Individuals over US Census Tracts for Nashville, TN}
    \includegraphics[width=.5\linewidth]{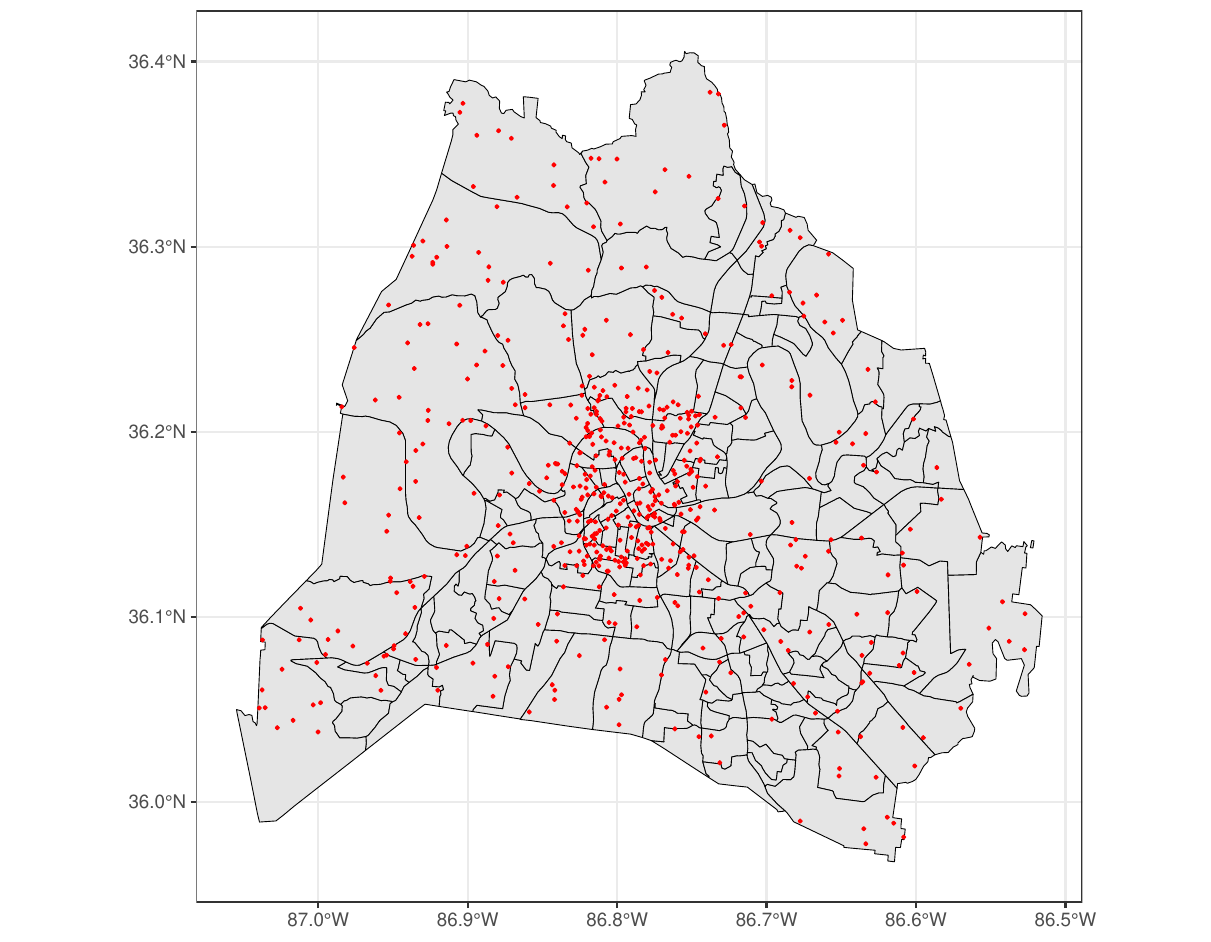}
    \caption{US Census tracts with simulated people experiencing homelessness locations.}
    \label{fig:tact_sim}
\end{figure}

\section{Implementation Discussion}

Our question was, ``how can a respondent-driven sample be deployed in a real-world setting to improve the PIT count?" Here, the authors propose the following strategy, built on our simulation work, known limitations of the HMIS database, and key concepts from the RDS literature.

The authors recommend establishing the spatial nature of the Continuum of Care in question, i.e., defining the geographic region and subregions where we will attempt to count the number of unsheltered people experiencing homelessness. For example, US Census-based tracts or blocks can provide a quality spatial sampling frame for the RDS. See, for example, Figure~\ref{fig:tact_sim}. Enumerators would ideally select sufficient locations to ensure good spatial coverage of the CoC (ideally, one would use past information on the locations of people experiencing homelessness and US Census-based units such as tracts; see, for example, Figure~\ref{fig:tact_sim}). The hubs would need to be fielded until a sufficient sample was obtained. One way to decide the necessary sample size would be to calculate a ``power analysis" that could use a simulated network, such as that developed in Section 9. The standard RDS procedure of setting up hubs over the CoC can be further refined by following the combined spatial/network sampling approach of \cite{felix2004combining}.

 \section{RDS for Estimating the Total Number of People Experiencing Homelessness in King County, WA}
In 2022, King County Regional Homelessness Authority (KCRHA), the CoC for King County, WA, received a waiver from HUD to conduct the Unsheltered PIT count through the proposed RDS strategy in this article.\footnote{The team at the University of Washington received an IRB Waiver (IRB ID: STUDY00015310), as data collection would be conducted regardless of University involvement.} King County (WA-500 CoC)\footnote{KCRHA was not yet established at this point as an organization.} had received a waiver not to have to conduct the 2021 unsheltered PIT count due to COVID-19\footnote{Sheltered PIT count was still required.} and had originally planned not to conduct the 2022 unsheltered PIT count but to conduct a large-scale qualitative study instead, where KCRHA would be doing a large-scale needs assessment of people experiencing homeless after the peak of COVID-19 pandemic. However, HUD mandates an unsheltered PIT to be completed at least bi-annually. Due to the waiver in 2021, 2022 was a required year for all CoCs that did not complete the unsheltered count 2021. KCRHA put out a call for proposals for alternative PIT count methods, which would allow for the originally planned qualitative study, and through this process, decided to propose the RDS method for both the qualitative and quantitative study with support from the team at the University of Washington. 

\begin{figure}[H]
    \centering
    \textbf{King County Continuum of Care (Seattle Metro)
RDS Unsheltered PIT Hubs, 2022}
    \includegraphics[width=.9\linewidth]{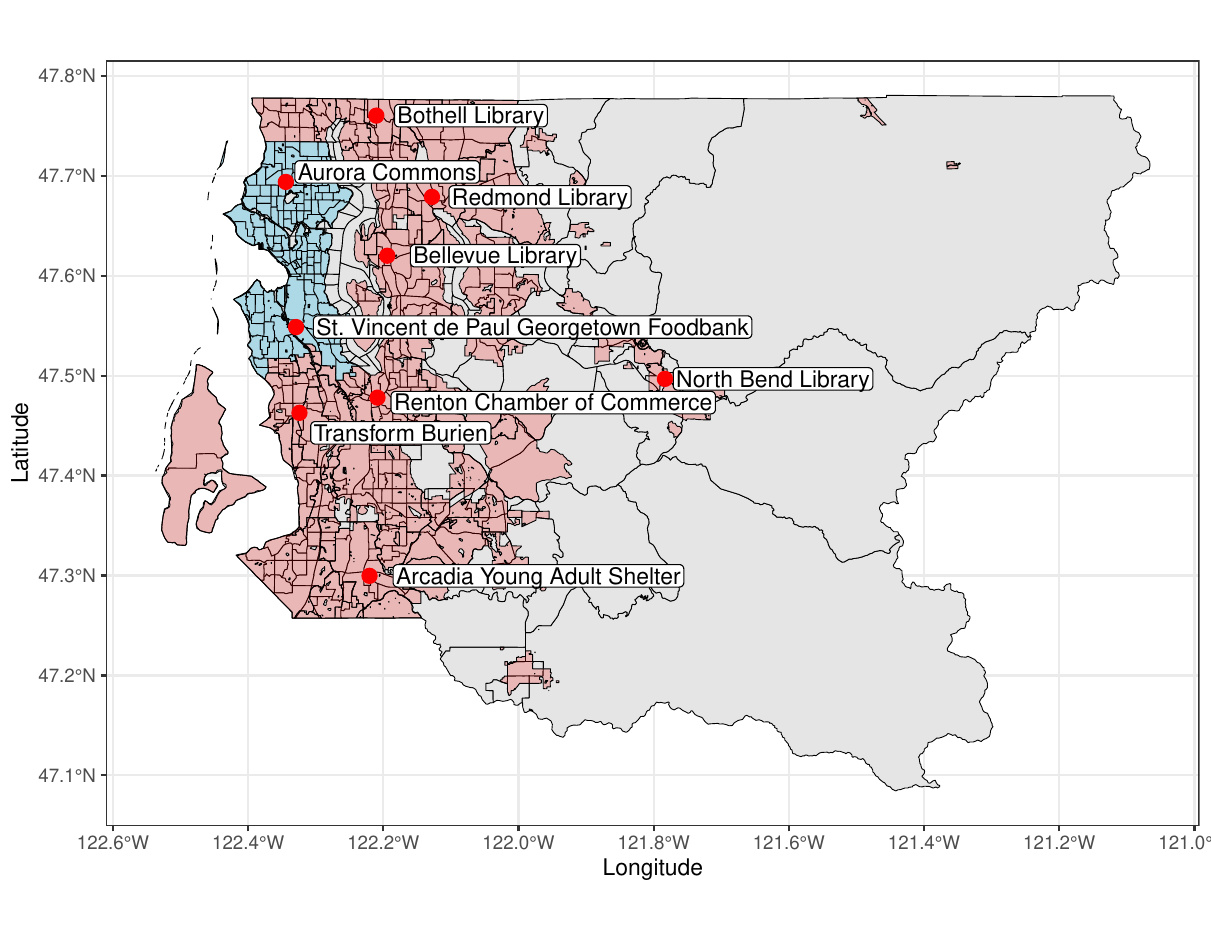}
    \caption{Nine hub locations used for the 2022 PIT count in King County, WA. US Census tracts are colored in grey, Seattle City US Census tracts are shaded in blue, and incorporated, and unincorporated urban areas are in red. The red dot is a hub location labeled with a named location.}
    \label{fig:rdskchub}
\end{figure}

KCRHA selected nine hub locations according to subregion based on previous PIT Counts feedback from people with lived experience. Where there are large populations of people living unsheltered (see Figure~\ref{fig:rdskchub}), they conducted data collection from March 9th to April 6th, 2022. All data is available on the HUD website \citep{HUDPIT2023}. KCRHA used volunteers from the local Lived Experience Coalition (LEC), which describes themselves as a ``largely BIPOC led"\footnote{\url{https://wearelec.org/}} activist organization with strong ties to the community of people experiencing homelessness, to conduct all interviews (qualitative and quantitative). A \$25.00 incentive was provided for all completed surveys, and three tickets (sometimes called coupons) with bus tickets were provided for recruitment. Seeds were obtained through the help of local service providers and appeared to be effectively random (removal of seeds from the estimation just produced more variance, but no noticeable effects on the mean estimate). The RDS sample obtained 671 interviews over the nine hub locations (see Table~\ref{tab:hublocsamp} for the breakdown). Overall, the largest chain was eight waves, resulting in an estimated 7,685 total unsheltered people experiencing homelessness using the estimator in Section 6.1. Standard sensitivity tests were performed, such as removing seeds and early waves of recruits (e.g., burn-in) and obtaining a target sample size (at least 600 respondents) \citep{gile2015diagnostics}. 

\begin{table}[H]
\centering
\textbf{Number of Respondents per Hub for King County, WA 2022 PIT Count}\\
\begin{tabular}{rp{6cm}r}
  \hline
 & HUB location & n \\ 
  \hline
1 & Arcadia Young Adult Shelter  & 104 \\ 
  2 & Aurora Commons & 172 \\ 
  3 & Bellevue Library &  63 \\ 
  4 & Bothell Library &   7 \\ 
  5 & Transform Burien &  36 \\ 
  6 & St. Vincent de Paul Georgetown Foodbank & 120 \\ 
  7 & North Bend Library &  15 \\ 
  8 & Redmond Library &  35 \\ 
  9 & Renton Chamber of Commerce  & 119 \\ 
   \hline
\end{tabular}
    \caption{Nine hub locations used for the 2022 PIT count in King County, WA, and the number of people surveyed at each hub.}
    \label{tab:hublocsamp}
\end{table}

\subsection{Comparing the Visual PIT to the RDS PIT in King County, WA}
The unsheltered PIT count was not conducted in 2021 in King County, WA, due to COVID-19, and the only unsheltered PIT count conducted in 2022 was the RDS unsheltered PIT count discussed in the last section. To build a comparative case study, we constructed an autoregressive integrated moving average (ARIMA; \cite{hyndman2008automatic}) model to impute the 2021 and 2022 unsheltered PIT count. We use the R package `forecast'\citep{hyndman2008automatic} to find the best fitting model over the 14-year period (2007-2020) of unsheltered PIT data in King County, WA \citep{HUDPIT2023}. We employ the AIC model fit criterion to find the best-fit model \citep{akaike1998information}. We find that an ARIMA(0,1,0) with a covariate of the shelter count (available in all years) is the best-fitting model (model parameters are available in Appendix A.1). In Table~\ref{tab:RDSvARIMA}, we can see the RDS estimate of 7,685 unsheltered people experiencing homelessness compared to 6,819 unsheltered people experiencing homelessness forecast of the visual PIT count. The statistical confidence interval highly overlaps, and the two estimates would not be statistically distinguishable. Overall, this is strong evidence that the two methods should be highly correlated, but there are some obvious advantages: (1) the demographic survey is conducted at the same time and on the same population as the unsheltered count compared to the historic PIT count model where the demographic survey is conducted at a later time and on a potentially different population \citep{allhome}; (2) the RDS estimate has a confidence interval and statistical uncertainty in its formulation that the visual census does not; and (3) there are clear strategies going forward to reduce estimate error. Further, this method allows for the people experiencing homelessness to engage in the process and express their own voices.

\begin{table}[H]
    \centering
\textbf{Forecasted Visual Unsheltered Count Versus RDS Estimate for King County, WA 2022}
\begin{table}[ht]
\centering
\begin{tabular}{rrr}
  \hline
 & RDS & ARIMA \\ 
  \hline
Point Estimate & 7,685 & 6,819 \\ 
  Lower Bound (95\% CI) & 6,816 & 5,277 \\ 
  Upper Bound (95\% CI) & 8,555 & 8,360 \\ 
   \hline
\end{tabular}
\end{table}
    \caption{RDS Estimate of the Total Unsheltered population experiencing homelessness in 2022 with statistical confidence interval (95\%) computed using the delta method \citep{oehlert1992note} compared to the best forecast of the visual PIT count conducted by an ARIMA(0,1,0) with covariate of the shelter count (available in all years) and selected by AIC criterion \citep{akaike1998information}.}
    \label{tab:RDSvARIMA}
\end{table}

\section{Discussion}

HUD delegates to CoCs the responsibility for counting the number of people in outdoor and sheltered homeless situations. Although soundly critiqued, the ``one night" crude census approach to estimating unhoused populations has become entrenched as usual. Enumeration strategies have been limited by a lack of resourcing and a desire to ensure a method that will not impact CoC funding. It is important to remember that most CoCs are small, with minimal staff. Counting is essential because we cannot solve problems we cannot describe. 


Recently, RAND conducted a repeated measure approach as a sensitivity analysis of the standard census-style PIT count, finding in the three Los Angeles neighborhoods they studied that unsheltered homelessness increased from 13\% to 32\% between 2021 and 2022 based on the Los Angeles Longitudinal Enumeration and Demographic Survey (LEADS) as the PIT count was not conducted in 2020 do to COVID-19 exemption by HUD \citep{ward2022recent}. There is some concern that some of this difference is due to \emph{sampling variability}, which would also be a concern for an RDS sample; however, Figure~\ref{fig:rds_bias} demonstrates we do not expect much statistical bias in the main measure and thus would not expect sampling variability to be a major concern. 

For much of the policy world and the general public, the only measure used will be the central tendency, so having an estimate low in statistical bias is paramount---and, we would argue, the most important feature of an estimator for policy uses. While we acknowledge the RDS estimator does provide more variance than we would like (this can be improved; see the discussion at the end of this section), the simulation and comparison studies show that we expect this single most used number, the mean estimate, to be of high-quality (see Figure~\ref{fig:rds_bias}) and also to provide some basic guidance (e.g., actual statistical bounds) to remind policymakers that it is a statistical estimate and that some caution around the interpretation of the number is warranted. It also provides for logistical purposes a lower and upper bound in expectation of the support one is likely to need for the unsheltered population. 

There are several ways demographic information on people experiencing homelessness could be acquired. Still, a solution that provides a good framework for a sampling-based approach with uncertainty bounds and a straightforward way to give the community a voice in their experience is a particularly compelling strategy. Here, we introduced one such strategy, built on the large body of work in public health for measuring hard-to-reach populations. We demonstrated in Davidson, TN, and King County, WA, that if we were to conduct respondent-driven sampling to obtain an estimate of the percent of unsheltered people experiencing homelessness, we could extrapolate to estimate the count of the unsheltered population of people experiencing homelessness by leveraging the known population of people using emergency shelters at the same time. The RDS framework provides an ethical approach, i.e., giving people the chance to volunteer to be in the count or not without being contacted by a researcher, and does not require invasion of someone's sleeping space by having enumerators prowl around with flashlights in the night hoping to catch sight of people "sleeping rough," as is done with PIT counts. 

Further, it provides a principled way to navigate the social network for sampling without a sampling frame or framework for conducting a power analysis (i.e., knowing the level of precision we need for policy analysis for the count of unsheltered people experiencing homelessness). Further, surveys cover people using emergency shelters and those on the streets,  providing clear quality assurance by leveraging the proportion estimates from the RDS method against the known administrative proportions (see Section 6: Empirical Example: Nashville, TN). Last, it provides a framework for adding additional qualitative surveys. When conducting a qualitative assessment of people experiencing homelessness, it's necessary to reach enough people to give a meaningful sense of the critical dimensions; the RDS framework also offers the opportunity to recruit people into a qualitative survey. This dual use is essential to CoCs that require a person count and a regular demographic report to HUD. After establishing the viability of RDS for PIT estimates, improvements on the data collection methods can reduce the variance of the estimators, for example, simulated sampling approaches \citep{verdery2015respondent} or linked egonet estimator \citep{lu2013linked} and simulation-based estimates \citep{gile2011improved}. Last, in cases where RDS is employed and other methods such as multiple-list \citep{weare2019counting}, scale-up methods \citep{feehan2016generalizing,killworth1990estimating}, or online surveys \citep{feehan2019using} are also employed, future work could be integrated with \cite{wesson2018bayesian}, which provides a Bayesian framework for synthesizing multiple estimators for hard-to-reach populations.

\section*{Acknowledgments}

\paragraph{Funding} Partial support for this research came from a Eunice Kennedy Shriver National Institute of Child Health and Human Development research infrastructure grant, P2C HD042828, to the Center for Studies in Demography \& Ecology at the University of Washington; NSF CAREER Grant \#SES-2142964; UW Population Health Initiative Tier 2 Grant. The content is solely the authors' responsibility and does not necessarily represent the official NIH or National NSF views. 

\paragraph{Partnership} King County Regional Homelessness Authority partnered with a team of researchers at the University of Washington led by Zack W. Almquist to conduct the unsheltered 2022 PIT RDS survey design and analysis. The King County Regional Homelessness Authority, as the Community of Care (CoC) lead, conducted and funded the survey project as part of their 2022 Point-in-Time count.

\paragraph{Conflict of Interest} Zack Almquist, Ashley Hazel, Mary-Catherine Anderson, Larisa Ozeryansky, and Amy Hagopian have no conflict of interest to declare. Owen Kajfasz, Janelle Rothfolk, and Claire Guilmette, as King County Regional Homelessness Authority employees, acknowledge the importance of transparency and accountability in scientific research and peer review. The King County Regional Homelessness Authority employed Kajfasz, Rothfolk, and Guilmette during the 2022 Point-in-Time count. The King County Regional Homelessness Authority, in its role as the Community of Care (CoC) lead, both conducted and provided funding for the King County 2022 RDS survey, which was carried out as a requirement of the HUD-mandated biannual 2022 Point-in-Time count. \textbf{Financial Interests}: Kajfasz, Rothfolk, and Guilmette declare that they have no financial interests, such as stocks, patents, or research grants, that may be perceived as affecting their objectivity in the peer review process. \textbf{Organizational Interests}: The King County Regional Homelessness Authority has a stake in addressing homelessness in King County and may be impacted by the outcomes of this survey project. As the Community of Care (CoC) lead, the King County Regional Homelessness Authority must complete a bi-annual unsheltered Point-in-Time count by both Federal and State statutes as a requirement for continued funding from the Department of Housing and Urban Development. As such, they have a stake in identifying valid methods of enumeration of people experiencing unsheltered homelessness. 

\bibliographystyle{ECA_jasa}
\bibliography{peh}

\appendix
\section{Appendix}

\subsection{ARIMA Model for Section 11.1}

\begin{table}[H] \centering 
\textbf{RDS versus Forecast of Visual PIT Count}
\begin{tabular}{@{\extracolsep{5pt}}lc} 
\\[-1.8ex]\hline 
\hline \\[-1.8ex] 
 & \multicolumn{1}{c}{King County Unsheltered Count} \\ 
\cline{2-2} 
\\[-1.8ex] & Coefficient (Standard Error) \\ 
\hline \\[-1.8ex] 
 drift & 306.69$^{**}$ (144.94) \\ 
  log(Shelter Count) & $-$7,580.84$^{*}$ (4,573.39) \\ 
 \hline \\[-1.8ex] 
Observations & 13 \\ 
Log Likelihood & $-$99.52 \\ 
$\sigma^{2}$ & 309,249.90 \\ 
Akaike Inf. Crit. & 205.05 \\ 
\hline 
\hline \\[-1.8ex] 
\textit{Note:}  & \multicolumn{1}{r}{$^{*}$p$<$0.1; $^{**}$p$<$0.05; $^{***}$p$<$0.01} \\ 
\end{tabular} 
  \caption{Fitting an ARIMA(0,1,0) regression to King County Unsheltered PIT data, 2007-2020, to forecast the 2022 visual PIT count. The best-fit model selected by AIC.} 
  \label{tab:arimamodel} 
\end{table}


\end{document}